\documentstyle[12pt]{article}

\input epsf.sty

\newcommand{\be}[1]{
\begin{eqnarray}\label{#1}}
\newcommand{\ee}{\end{eqnarray}}
\newcommand{\ci}[1]{\cite{#1}}
\newcommand{\re}[1]{(\ref{#1})}
\renewcommand{\thefootnote}{\fnsymbol{footnote}}

\def\Gtilde{\tilde{G}}

\newcommand{\dxi}{\frac{\partial}{\partial \xi}}

\newcommand{\dx}{\frac{\partial}{\partial x}}
\newcommand{\du}{\frac{\partial}{\partial u}}
\newcommand{\eps}{\varepsilon}

 \def\Tr{\mbox{Tr}}

\newcommand{\insertfig}[2]{\mbox{\epsfxsize=#1cm \epsfbox{#2.eps}}}

\begin{document}
\renewcommand{\thefootnote}{\fnsymbol{footnote}}
\begin{flushright}
\begin{tabular}{l}
 TPR-01-07
\end{tabular}
\end{flushright}
\begin{center}
{\bf\Large  
On hard electroproduction of mesons: helicity flip
 amplitudes with tensor gluon contributions
}

\vspace{0.5cm} N. Kivel$^{a}$\footnote{on leave of absence from 
St.Petersburg
Nuclear Physics Institute, 188350, Gatchina, Russia }, 

\begin{center}
{\em $^a$ Institut f\"ur Theoretische Physik, Universit\"at
Regensburg \\ D-93040 Regensburg, Germany}
\\[10mm]

\end{center}

\end{center}

\begin{abstract}

We consider  hard electroproduction of mesons
at twist-3 level and focus our attention on 
helicity flip amplitudes corresponding to the production of
transversely polarized mesons by transverse photons.
We demonstrate that helicity-flip amplitude for the transversely polarized
particles are closely related to the contributions of the matrix 
elements of the tensor gluon operator which  can serve as pure gluonic 
probe of the hadrons.  Using  standard factorisation approach
and   Wanzura-Wilczek approximation for the twist-3 matrix elements
we find that  such amplitudes do not have divergencies from the endpoint
regions  which appear in the other twist-3 amplitudes.

\end{abstract}

\newpage
\section*{\normalsize \bf Introduction}

Hard exclusive reactions provide a unique opportunity to study the partonic 
structure of the hadrons, 
see e.g. recent reviews \ci{GPV,RadR} and references there in. 
In recent years exclusive electroproduction of mesons 
from nucleons has become a topic of broad interest. At higher energies a large 
amount of data has become available from experiments at DESY (HERA, HERMES)
\ci{Zeus, herm1, herm2}. Further measurements are expected in future at DESY
and CEBAF. From the theoretical side the progress in description of exclusive
electroproduction has been possible due to factorisation theorem \ci{CFS}. 
It states that
the underlying photon-parton sub-processes are dominated by longitudinally
polarized photons and large photon virtualities, $Q^2\gg \Lambda^2_{QCD}$ and
hence can be calculated perturbatively.  Using  such OPE approach productions 
of mesons by longitudinal photons have been studied in series of papers  
\ci{Brodsky,MPR,MPW, Radyushkin,Van1,Van2,LDP,Fran}.  

The situation with description of interactions of transversely 
polarized photons is less trivial.  According to analysis carried out in
\ci{CFS} such amplitudes are higher twists effects, i.e. suppressed as $1/Q$
compared to the longitudinal photon. At the same time experimental measurements 
 show clearly that helicity conserving amplitudes with transverse photon
provide large contribution to observables at moderate values of $Q^2$. 
Hence one must  deal with higher twists effects to understand the 
dynamics of the hard exclusive electroproduction. Unfortunately,
 straightforward application of pQCD is not possible  because the
factorisation is violated for the meson production by a transverse photon 
\ci{CFS,Brodsky,Rad,MP}.

The violation of the factorisation for the transverse case is closely related to 
the so-called  end-point or soft contributions \ci{CFS,Brodsky,Rad}.
Using terminology of reduced diagrams \ci{CFS, Rad} the   
OPE and endpoint contributions for  hard electroproduction process 
are shown in Fig.~1 and Fig.~2 respectively.
In case of the  transverse virtual photon or transverse outgoing 
meson the endpoint  
regime  contributes to the same accuracy  $1/Q^2$ 
as  short distance configuration  Fig.~1 \ci{CFS,Brodsky,Rad}.  
Practically, the  divergencies in the convolution integrals of the
OPE-like configuration  can be considered as a test for the 
possible contributions associated with endpoint  diagram Fig.~2.
A direct calculation carried out for the transverse
$\rho$-meson \ci{MP} supports this observation. On the other hand,
if the endpoint divergencies are absent in the convolution integrals 
one can hope  that soft configurations give sub-leading  contribution to 
an amplitude of this process.
      
In present paper we shall study helicity-flip amplitudes 
in the hard exclusive meson electroproduction.  
A typical feature of such processes is a contributions
of the tensor gluon operators. Recently helicity-flip  amplitudes 
have been the subject of theoretical research  in DVCS \ci{JiH,BelMu,Diehl}, 
fragmentation \ci{Ter} and
 meson production in  $\gamma^*\gamma$ fusion \ci{BK, KMP}.     
We shall demonstrate that  
helicity-flip amplitudes with tensor gluons 
in the exclusive meson electroproduction
do not possess the endpoint divergencies and 
hence one can hope that  these amplitudes can 
be described in the framework of QCD factorisation.  
For simplicity we shall consider scalar pion target, the generalization
to the nucleon target is straightforward. 

The remainder of this paper is organized as follows: in the next section we
discuss the amplitude for the hard electroproduction of tensor mesons.
The following section is devoted to
discussion of the helicity-double-flip amplitude for the case of  
$\rho$-meson electroproduction. Finally, we conclude. Some technical details
of present calculation are summarized in the Appendix.

\section*{\normalsize \bf Hard electroproduction  of 
spin-2 meson  with tensor polarization }

The amplitude of the  process
\be{procf}
\gamma^*(q)+\pi(p)\to f_2(Q') +\pi(p') \, ,
\ee
is defined in
terms of the  matrix element of the 
electromagnetic current~:
\be{T:def}
T^{\mu}=\langle \pi(p'),f_2(Q',\lambda)|J_{\rm
e.m.}^\mu (0)|\pi(p)\rangle\, ,
\ee
where  index $\lambda$  corresponds to the helicity of the tensor meson.

We shall consider the Bjorken limit, where collision energy $W$
and virtuality of the photon are large: $W\to \infty$, 
$-q^2=Q^2\to\infty$, but their ratio $Q^2/W$ is fixed. 
Let $p, p'$ and $q$ denote the
momenta of the initial and final pions and photon, respectively.
We introduce also light-cone vectors $n,\, n^*$ such that 
\be{nnstar} n\cdot n=0, \, n^*\cdot
n^*=0, n\cdot n^*=1. 
\ee  
We shall work in the reference frame where the average nucleon
momenta $P =\frac12(p+p')$ and the virtual photon momentum  $q$
are collinear along the z-axis and have opposite directions.
To the twist-3 accuracy such a choice of the frame results in the 
following decomposition for the  momenta:
\be{lce} 
P&=&\frac12(p+p')=n^*, \quad \Delta = p'-p =-2\xi
P+\Delta_\perp, \, \nonumber 
\\[4mm] 
q&=&-2\xi
P+\frac{Q^2}{4\xi}n, \quad
Q'=q-\Delta=q'+\frac{m^2}{2(Pq')}n^*-\Delta_\perp\, , \\
Q'^2 &=& m^2,\, \quad q'= \frac{Q^2}{4\xi}n\, .
\ee
where $m=1270$~MeV is the $f_2$-meson mass.

We also define the transverse metric and
antisymmetric transverse epsilon tensors
\footnote{The Levi-Civita
tensor $\epsilon_{\mu \nu \alpha\beta}$ is defined as the totally
antisymmetric tensor with $\epsilon_{0123}=1$ }:
\be{gt} g^{\mu
\nu}_\perp = g^{\mu \nu}- n^\mu n^{*\, \nu}-n^\nu n^{*\, \mu},
\quad \epsilon^\perp_{\mu \nu}= \epsilon_{\mu \nu
\alpha\beta}n^\alpha n^{*\,\beta}\, . 
\ee 
%
%
%

We start from  short discussion of  the amplitude in the leading
twist approximation.  This amplitude describes the scattering of
longitudinal photon and production of longitudinal meson. 
The factorisation theorem \ci{CFS} states that such amplitude 
 is dominated  by OPE-configuration shown in Fig.~1 and can be written 
in the form of integral convolution of a hard and soft collinear parts.  
The hard part  is given by scattering amplitude of partonic subprocess
and can be calculated in pQCD. The soft blocks are represented in terms of
 matrix elements of the light-cone operators. 
Typical diagrams contributing to our process are shown in Fig.~4 and Fig.~5. 

Let us define twist-2 skewed parton distributions (SPD's) and
distribution amplitudes (DA's) which parameterize soft matrix elements
in the  expression for the leading twist amplitude.

Twist-2 SPD of the pion is given by 
\footnote{The gauge link between points on the 
light-cone is not shown but always assumed.}  
\be{Htw2}
\langle p'| \bar \psi\biggl(\frac{\lambda}{2}n
\biggr) \gamma^\mu
\psi\biggl(-\frac{\lambda}{2}n\biggr)|p\rangle&=&
P^\mu \int_{-1}^1 dx e^{i x \lambda}H(x,\xi)\, ,
\ee
where, for the sake of simplicity,  
we do not write explicitly the $t$-dependence of the SPD, denoting
$H(x,\xi)\equiv H(x,\xi,t)$. Notice that this  SPD is symmetric under the 
interchange of $x\leftrightarrow -x$:
\be{Hsym}
H(-x,\xi)=H(x,\xi)
\ee

Twist-2 DA's for the tensor meson have been introduced in \ci{BK}. 
Taking into account the  kinematics \re{lce} we obtain:
\be{phi}
 \langle f_2(Q',\lambda)|\bar \psi(n^*)\gamma_\mu \psi(-n^*)|0\rangle
 =
 f_q m^2 q'_\mu \frac{e^{(\lambda)}_{\alpha\beta}n^{*\alpha}n^{*\beta}}
{(q'n^*)^2}
  \int\limits_{-1}^1 \! du \,e^{iu(q'n^*)} \phi_q(u)\, ,
\ee
where $\bar \psi \psi = \bar u u+\bar d d$ and the symmetric and traceless
 polarization tensor $e^{(\lambda)}_{\alpha\beta}$ 
 satisfies the condition 
$e^{(\lambda)}_{\alpha\beta} Q'^\beta=0$.
Polarization sums can be calculated using 
\be{polarization}
\sum_\lambda e^{(\lambda)}_{\mu\nu}\left(e^{(\lambda)}_{\rho\sigma}\right)^\ast
  = \frac12 M_{\mu\rho} M_{\nu\sigma}+\frac12 M_{\mu\sigma} M_{\nu\rho}
  -\frac13 M_{\mu\nu} M_{\rho\sigma}\,,
\ee
where $M_{\mu\nu} = g_{\mu\nu} - Q'_\mu Q'_\nu/m^2$ and 
the normalization is such that 
$e^{(\lambda)}_{\mu\nu}\big(e^{(\lambda')}_{\mu\nu}\big)^\ast= 
\delta_{\lambda\lambda'}$. 

In addition, there exists  leading twist 
gluon distribution amplitude:
\be{GS}
\langle f_2(Q',\lambda)|
S_{\mu\nu}G^a_{\xi\mu}(n^*)G^a_{\xi\nu}(-n^*)|0\rangle
 &=&
 f^S_g e^{(\lambda)}_{\mu\nu} \int\limits_{-1}^1 \! du\,
e^{iu(q'n^*)} \phi_g^S(u)\, .
\ee
Here $S_{\mu\nu}$ stands for the  symmetrisation in the two indices and 
removal of the trace: 
$ S_{\mu\nu}{\cal O}_{\mu\nu} = \frac12\, {\cal O}_{\mu\nu} +
\frac12\,{\cal O}_{\nu\mu}
 -\frac14\, g_{\mu\nu}{\cal O}_{\xi\xi}$.

The distribution amplitudes $\phi_q(u) = -\phi_q(-u)$ and 
$\phi_g^S(-u)=\phi_g^S(u)$  are the leading twist-2 
distribution amplitudes for the tensor mesons with helicity $\lambda=0$.
The constants $f_q$ and $f^S_g$ are defined as the matrix elements of the 
local operators, for details see  \ci{BK}.

In terms of these functions  expression for leading twist amplitude reads:
\be{LT}
T^\mu_{\gamma_L\to f_2(0)}&=&\frac{\left(2\xi P+q'\right)^\mu}{(Pq')}
\frac{e^{(\lambda)}_{\alpha\beta}P^\alpha P^\beta}{(Pq')} 
\frac{m^2}{(Pq')}(4\pi e\alpha_S) \frac{1}{4N_c}\frac1\xi
\nonumber \\[4mm] 
&&\times\int_{-1}^1 dx H(x,\xi)C^-(x,\xi)\int_{-1}^1 du \frac1{1-u^2}
\left\{\frac12 C_F f_q u\phi_q(u)+f^S_g\phi^S_g(u)\right\}\, ,
\ee
where 
\be{Cpm}
C^{\pm}(x,\xi)= \frac1{x-\xi+i0}\pm \frac1{x+\xi-i0}\, .
\ee

The flavor dependence in Eq.\re{LT} can be restored easily by substitution:
\be{flavor}
H(x,\xi)&\to& e_u H_u(x,\xi)+e_d H_d(x,\xi)\, ,
\ee
where
$e_f$ is charge of a quark of flavor $f=u,d$ in units of the 
proton charge ($e_u=2/3, e_d=-1/3$).

All convolution integrals in \re{LT} are well defined as it should 
be according to factorisation theorem. It is obvious, that amplitude
\re{LT} can be considered as a  particular limit  of more general situation
of production of two pions from longitudinal photon. Such process
have been considered in \ci{LDP} and our result \re{LT} is in 
agreement with that presented in \ci{LDP}.

At the twist three level
\footnote{We adopt here kinematical definition of twist, i.e.
terms suppressed by $1/Q^2$ are of twist-3.} 
there exist three  amplitudes with helicity flip:
$\gamma_\perp(\pm 1)\pi\to f_2(0)\pi$,
$\gamma_L(0)\pi\to f_2(\pm1)\pi$ and 
$\gamma_\perp(\pm 1)\pi\to f_2(\pm2)\pi$.
In what follows we shall consider  helicity flip amplitude describing 
transition $\gamma_\perp(\pm 1)\pi\to f_2(\pm2)\pi$ because 
tensor gluon contribution can appear only in this case. Let us note
that diagrams like in Fig.~5 can not contribute to the twist-3 accuracy to
such amplitude because meson state with $\lambda=\pm 2$ constructed from two
quarks in P-wave is suppressed by extra power of large $Q^2$ in comparison
with gluons carrying equal helicities.  

The space time development of partonic process with such gluons is shown 
in  Fig.3. For simplicity  we refer to the region 
where all momentum fractions are positive and hence the partonic subprocess is
\be{parton}
\gamma^*+q\to gg +q
\ee         
As one can see from that picture, in the pure collinear kinematics
the projections of the angular momentum $J_z$ in initial and final states 
are not balanced . To match them we must
consider diagrams with emission of the transverse gluons 
(genuine twist-3 contributions) 
or  include kinematical effects due to orbital motion of the  quarks.
In what follows  we  neglect for simplicity the genuine 
twist-3 contributions and consider only kinematical or 
Wandzura-Wilczek (WW) contributions to the twist-3 matrix elements. 

In   our calculations we shall use  following matrix elements. The
tensor meson with  helicities $\lambda=\pm 2$ is created by two collinear 
gluons with equal  helicities, see Fig.~3. Corresponding twist-2 matrix element
have already been  considered in \ci{BK}:
\be{GT}
\langle f_2(Q',\lambda)|
S_{\mu\nu}G^a_{n^*\mu}(n^*)G^a_{n^* \nu}(-n^*)|0\rangle
 &=&
 f^T_g e^{(\lambda)}_{\mu\nu} (q'n^*)^2 \int\limits_{-1}^1 \! du\,
e^{iu(q'n^*)} \phi_g^T(u) \,,
\ee          
where we used notation $G_{n^*\mu} = n^*_\xi G_{\xi\mu}$. 

The distribution amplitude $\phi_g^T(u)$ is 
 symmetric with respect to the interchange 
of $u\leftrightarrow -u$ and describes the momentum fraction distribution
of the two gluons in the $f_2$-meson. 
The asymptotic distribution at large scales is equal to 
\be{phi2as}
   \phi_g^{T,{\rm as}}(u) =  \frac{15}{16} (1-u^2)^2\,. 
\ee  
   The constant  $f^T_g$ is renormalized multiplicatively
\cite{JiH,LFBK85,Eric}:
\be{GTscale}
f^T_g(Q^2)& = & f^T_g(\mu^2) L^{-1+6N_c/\beta_0}\,, 
\ee
where  $L=\alpha_s(Q^2)/\alpha_s(\mu^2)$, $\beta_0 = 11/3 N_c -2/3 n_f$.

Expressions for the pion matrix elements 
must be written up to twist-3 accuracy. 
Corresponding  results have been obtained in \ci{BM, KPST,RW1,An}. 
We shall follow notation adopted in \ci{ KPST,ATP}:
\be{V}
\int \frac{d\lambda}{2\pi} e^{i\lambda x}
\langle p'| \bar \psi\biggl(-\frac{\lambda}{2}n
\biggr) \gamma^\mu
\psi\biggl(\frac{\lambda}{2}n\biggr)|p\rangle&=&
P^\mu H(x,\xi) + \Delta^\mu_\perp H_3(x,\xi)\, , \\
\label{A}
\int \frac{d\lambda}{2\pi} e^{i\lambda x }
\langle p'| \bar \psi\biggl(-\frac{\lambda}{2}n
\biggr) \gamma_\mu \gamma_5
\psi\biggl(\frac{\lambda}{2}n\biggr)|p\rangle&=&
i\varepsilon_{\mu \alpha \beta \delta}\Delta^{\alpha}P^{\beta}
n^{\delta}H_A(x,\xi) \, .
\ee

In the WW-approximation twist-3 SPD's $ H_3(x,\xi)$ and $ H_A(x,\xi)$  
can be rewritten in terms of twist-2 function $H(x,\xi)$. Corresponding
expressions read:
\be{WW3}
H_3^{WW}(x,\xi)&=& -\frac12\int_{-1}^1 du W_+(x,u,\xi)\dxi H(u,\xi)
-\frac12\int_{-1}^1 du W_-(x,u,\xi)\du H(u,\xi)\, ,
\\[4mm]\label{WWA}
H_A^{WW}(x,\xi)&=& \frac1{2\xi}\int_{-1}^1 du 
W_-(x,u,\xi)\left(\xi\dxi+u\du\right)H(u,\xi)\, ,
\ee  
where WW-kernels read
\be{Wpm}
W_{\pm}(x,u,\xi)&=& \frac12\biggl\{
\theta(x>\xi)\frac{\theta(u>x)}{u-\xi}-
\theta(x<\xi)\frac{\theta(u<x)}{u-\xi} \biggl\} \nonumber
\\[4mm]&&\mskip-10mu \pm\frac12\biggl\{
\theta(x>-\xi)\frac{\theta(u>x)}{u+\xi}-
\theta(x<-\xi)\frac{\theta(u<x)}{u+\xi} \biggl\}.
\ee

Typical diagrams  contributing to the amplitude 
$\gamma_\perp(\pm 1)\pi\to f_2(\pm2)\pi$ at twist-3 level 
are depicted in the Fig.~4. Using 
matrix elements \re{GT}, \re{V} and \re{A} we obtained:
\be{Tgf}
T^\mu_{\gamma_\perp(\pm 1)\to f_2(\pm2)}&=&
e^{(\mu_\perp \nu_\perp)}\Delta_{\perp \nu}T_2(Q^2,\xi),
\\[4mm]
\label{Tgf2}
T_2(Q^2,\xi)&=&
 \frac{f^T_g}{Q^2}
(4\pi e\alpha_S )\frac{8}{N_c}
\int_{-1}^1du \frac{\phi^T_g(u)}{(1-u^2)^2}
\\[4mm]
&&  \times \int_{-1}^1 dx\, C^-(x,\xi)\left\{
\left(x\dx+\xi\dxi\right)H(x,\xi)
-\xi H_3(x,\xi)-x H_A(x,\xi)
\right\}\, ,\nonumber 
\ee
where $e^{(\mu_\perp\nu_\perp)}$ is symmetric and traceless transverse
projection of the polarization vector $e^{(\lambda)}_{\mu\nu}$. 
The flavor dependence in Eq.\re{Tgf} can be restored by same substitution
as in \re{flavor}.

Consider the convolution integrals which appear in the RHS of \re{Tgf}.
The integral with meson distribution amplitude has now stronger endpoint
singularities in the coefficient function in comparison with twist-2 case \re{LT}. 
But we expect that gluon distribution amplitude  vanishes faster than $u$ as
$u\to \pm1$. At least asymptotical shape \re{phi2as} supports such expectations.
 Then  singularities at $u=\pm 1$ of the coefficient function is compensated 
by endpoint behavior of the gluonic wave function and the integral 
is convergent.  

The second integral, defining convolution with
pion SPD's is also well defined. To see this, 
one has to inspect properties of the convolution integrals which
appear in \re{Tgf} at the points $x=\pm \xi$.  Consider first
combination like $\left(x\dx+\xi\dxi\right)H(x,\xi)$. It has no 
discontinuities at the points  $x=\pm \xi$  \ci{KPV}. 
One can see this easily integrating by parts
and using that $H(x=\pm 1 ,\xi)=0$:
\be{1st}
\int_{-1}^1 du\, C^-(x,\xi)
\left(x\dx+\xi\dxi\right)H(x,\xi)=
\xi\dxi\int_{-1}^1 dx\, C^-(x,\xi) H(x,\xi).
\ee  
where RHS is well defined. 
The remaining part of the convolution integral with SPD in \re{Tgf} 
depends on the combination $\xi H_3(x,\xi)+x H_A(x,\xi)$ which
is free from discontinuities at the points $x=\pm \xi$  \ci{KPST,RW1,BMKS}. 
One can prove it in WW-approximation using \re{WW3} and \re{WWA}.

In general case, the factorisation for twist-3 amplitude for the 
meson production is broken as it have been discussed in \ci{CFS}.   
This is  closely  related to the fact, that in addition to the 
OPE-like diagram Fig.~1, there is contribution of the same order
from other  reduced diagram depicted in Fig.~2.
This is exactly the, so-called, endpoint contribution discussed in 
\ci{Brodsky,Rad}. 
For instance, the divergencies of the convolution integrals
which were found in twist-3 calculations for the exclusive  
production of transversely polarized vector mesons \ci{MP} can be 
considered as a signal that contributions from the regions 
depicted in  Fig.~2  must be taken into account. 
We expect that such situation takes place also for   other
twist-3 helicity amplitudes which were mentioned above, namely, 
$\gamma^*_L \pi\to f_2(\pm 1)\pi$, \, 
$\gamma^*_\perp(\pm 1) \pi\to f_2(0)\pi$. 

However,  we observed that such  divergencies are absent in 
the expression \re{Tgf2}. 
We suppose that 
it  could be related to a possibility that OPE-like   
contribution dominates at twist-3 accuracy
in such helicity-flip amplitude. To see it rigorously one has, of course, 
to prove that the possible  endpoint contributions 
corresponding to Fig.~2 is sub-leading. 
We shall not address  this question in details in the paper. 
Some naive arguments in support
of our speculation can be given on the  basis   of conservation of 
angular momentum. To get the spin projection $S_z=2$ the quark and 
antiquark have to be in a 
$P$-wave state while the gluons can be in $S$-wave. On the light-cone,
however, contribution of the orbital angular momentum is higher-twist and 
amplitude corresponding to  helicity state $\lambda=\pm 2$ is dominated 
at large photon virtualities by gluon contribution. 
Of course, such arguments can not guarantee the absence of the 
soft configurations  even if contributions from the 
OPE-like configurations Fig.~1 are free from the divergencies. 
In this paper we adopt the point of view which 
 states that if  factorisation is not violated, i.e. there are no
 divergencies in the convolution integrals, then    
possible non-OPE configurations are sub-leading. From this assumption
it follows that helicity flip amplitude $\gamma^*_\perp(\pm 1) 
\pi\to f_2(\pm2)\pi$ can be described in the framework of factorisation 
approach.


Since $f_2$ decays in two pions with the branching ratio
about 85\%, one natural possibility to measure  the amplitudes
of interest is via the hard exclusive two-pion electroproduction 
$\gamma +p \to p \pi\pi$. This
process   has  been considered recently at the
leading twist approximation in \ci{LDP}.
Let $k_1$  and $k_2$ be the momenta of the two pions 
in the final state and $k=(k_1-k_2)$ is their relative momentum. 
The form  factor of interest
$T_2(Q^2,\xi)$ \re{Tgf} is related to the 
 corresponding two-pion  amplitude 
$A^{\pm}_{\gamma(\pm 1)\to \pi\pi(\pm2)}$
\footnote{We use notation $\varepsilon_\mu(\pm)A^\mu\equiv A^{\pm}$,
where $\varepsilon_\mu (\pm)$ denotes the transverse 
polarization vector of the virtual photon} 
in the  region $(k_1+k_2)^2\sim m^2$ as
\be{twopion}
A^{\pm}_{\gamma(\pm 1)\to \pi\pi(\pm2)} =\varepsilon_\mu (\pm)
 (k^\mu_\perp k^\nu_\perp-\frac12g_\perp^{\mu\nu}k_\perp^2)
\Delta_{\perp\nu} 
\frac{g_{f_2\pi\pi}}{m^2}\frac{T_2(Q^2,\xi)}{m^2-(k_1+k_2)^2},       
\ee
where $g_{f_2\pi\pi}$ is the corresponding decay constant:
\be{fpipi}
\langle \pi(k_1)\pi(k_2)|f_2(P,\lambda)\rangle=\frac{g_{f_2\pi\pi}}{m}
e_{\alpha\beta}^{(\lambda)}k^\alpha k^\beta\, .
\ee
Hence 
$A^{\pm}_{\gamma(\pm 1)\to \pi\pi(\pm2)}$ can be separated using
its nontrivial dependence on the azimuthal angle 
$\phi_{\pi\pi}$ between lepton and decay pion planes.
\footnote{To avoid  confusion we
stress that ``pion plane'' is the plane which is defined by 
pions from decay $f_2\to \pi\pi$.}  
 Moreover, $A_{\pm}$ is symmetric under the interchange of the 
pion momenta (isoscalar state) and can be measured by the 
interference with the  contribution of the two-pion production in 
isovector state   that is antisymmetric to the interchange 
of the pion momenta: In the difference of cross sections 
$\sigma(k_1,k_2)-\sigma(k_2,k_1)$ only the interference term survives.
Some estimates of the interference for the leading twist cross 
sections have been presented in \ci{LDP,LDP2}.

\section*{\normalsize \bf Hard electroproduction  of 
$\rho$-meson with double-helicity-flip }

In this section we consider another possibility to get twist-3 amplitude 
with tensor gluon operator. It is possible in the process 
of $\rho$-meson production with double-helicity-flip :
$\gamma^*(\pm 1)\pi\to \rho(\mp 1)\pi$. Such amplitude is 
responsible for the violation of the s-channel helicity conservation (SCHC)
\ci{SCHC}. The effects of the  violation of the SCHC in electroproduction
have been considered in various models in \ci{Gol,Sch,Ivan}.
Our  approach is  different because it based on the application of OPE.  
In such approach double-helicity-flip production of the 
$\rho$-meson is   analogous, in some sense, to helicity flip amplitude in 
DVCS \ci{JiH, BelMu}. But in DVCS such amplitude is leading twist-2 
contribution. In case of electroproduction this amplitude 
 is non-leading twist-3 effect, i.e. behaves as $1/Q^2$ because $\rho$ meson
is a composite particle. 

The amplitude of the process
\be{proc}
\gamma^*(q)+\pi^\pm(p)\to \rho_0(Q') +\pi^\pm(p') \, ,
\ee
is defined in
terms of the  matrix element of the 
electromagnetic current~:
\be{M:def}
M^{\mu}=\langle \pi(p'),\rho(Q',\lambda)|J_{\rm
e.m.}^\mu (0)|\pi(p)\rangle\, ,
\ee
Decomposition of  momenta can be  obtained from  \re{lce} using
substitution $m\to m_\rho$ where $m_\rho$ is the $\rho$-meson mass.
  
Typical diagrams contributing to $\rho$-meson helicity flip amplitude are depicted
in Fig.~6. The space time development of the corresponding hard partonic subprocess 
is shown in the Fig.~7.  For simplicity, 
we consider only situation when all momentum fractions are positive. 
Again, we observe  that
for  pure collinear scattering process there is a difference 
of angular momentum projections between initial and final partonic states. 
Therefore, one may conclude that  amplitude of  this process is suppressed 
at least by $1/Q$ relative to twist-2 amplitude of pure longitudinal 
$\rho$-meson production $\gamma^*_L\pi\to \rho_L\pi$. To obtain the leading
 contribution to the helicity flip amplitude one must take into account 
the twist-3 contributions to the $\rho$-meson matrix elements. 
Corresponding distribution amplitudes have  been introduced in
\ci{BB, BBKT}. As in the previous case, in our calculations we neglect 
genuine twist-3 effects and
consider only corresponding WW-contributions . 
We define chiral even two-particle distribution amplitudes of the $\rho$-meson as
\be{Vrho}
\langle \rho(Q',\lambda)|
\bar \psi (n^*) \gamma_\mu
\psi(-n^*)|0\rangle&=& f_\rho m_\rho
q'_\mu \frac{(e^* n^*)}{(q'n^*)}\int_{-1}^1 du \phi(u)e^{i(q'n^*)u}
\nonumber \\[4mm]
&&+f_\rho m_\rho e_{\perp\mu}^{*} \int_{-1}^1 du G(u) e^{i(q'n^*)u}\, ,
\\[4mm]
\label{Arho}
\langle \rho(Q',\lambda)|
\bar \psi (n^*) \gamma_\mu\gamma_5
\psi(-n^*)|0\rangle &=& \frac{i}{2}f_\rho m_\rho \epsilon_{\mu \nu\alpha\beta}e_\perp^{*\nu}
\frac{q'^\alpha n^{*\beta}}{(q'n^*)}
\int_{-1}^1 du \Gtilde (u) e^{i(q'n^*)u}\, ,
\ee        
where  $f_\rho$ and $e^*_\mu$ denote the decay constant and
polarization vector of a $\rho$-meson and $\bar \psi \psi= \bar u u-\bar d d$. 
Twist-3 distributions amplitudes $G(u)$ and
$\Gtilde (u)$ in the WW-approximation are given by following relations:
\be{WWG}
G(u)&=&\frac12
\int_{-1}^u dw \frac{\phi (w)}{1-w}+ 
\frac12\int_{u}^1 dw \frac{\phi (w)}{1+w}  \, ,
\\[4mm]
\label{WWGt}
 \Gtilde(u)&=&
\int_{-1}^u dw \frac{\phi (w)}{1-w}- 
\int_{u}^1 dw \frac{\phi (w)}{1+w} \, , 
\ee 

 Gluon SPD is defined by tensor gluon operator  in the following way:
\be{GTSPD}
\langle p'| 
S_{\mu\nu}G^a_{n\mu}(\frac{\lambda}{2}n)G^a_{n\nu}(-\frac{\lambda}{2}n)
|p\rangle&=&
 S_{\mu\nu}{\Delta_{\perp \mu}\Delta_{\perp \nu}}/{f_\pi^2}
\int_{-1}^1 dx e^{i x \lambda}G^T(x,\xi)\, ,
\ee
where we introduced a factor  $1/f_\pi^2$  in order 
to have the dimensionless combination
${\Delta_{\perp \mu}\Delta_{\perp \nu}}/{f_\pi^2}$. 
The SPD $G^T(w,\xi)$ is symmetric to the interchange of $x\leftrightarrow -x$
and  at large scale $\mu\to \infty$ the leading contribution equals to
\be{SPDas}
G^T_{as}(x,\xi)=G^T(\mu)\frac{15}{16}\frac1{\xi^5}(x^2-\xi^2)^2\theta(|x|\le \xi)\, .
\ee 
The  dimensionless constant $G^T(\mu)$ is renormalized multiplicatively with the
same anomalous dimension as $f_g^T$ in \re{GTscale}.

A straightforward calculation of Feynman diagrams, Fig.~7, gives:
\be{Mrho}
M^{\mu}&=& \frac{f_\rho m_\rho}{Q^2}(4\pi e \alpha_S)
\frac{ S_{\mu\nu}\Delta_\perp^\mu\Delta_\perp^\nu}{f_\pi^2}e^{*}_\nu
\frac{2\xi^2}{N_c}
\nonumber \\[4mm]
&&\times\int_{-1}^1 du \frac{\phi(u)}{1-u^2} \int_{-1}^1 dx 
\frac{G^T(x,\xi)}{(x-\xi+i0)^2(x+\xi-i0)^2}
\ee

Let us first examine the properties of the convolution integrals. The twist-2 
$\rho$-meson distribution amplitude which enters into the answer can be taken 
close to its asymptotic shape:
\be{phirho}
\phi(u)=\frac34(1-u^2)\, .
\ee
Then one can see that corresponding convolution integral in \re{Mrho} 
 is well defined. It is 
interesting that possible contributions of twist-3 distributions amplitude $G(u)$
and $\Gtilde(u)$ cancel in the sum of all diagrams. We do not see special reasons 
for this, except that it can be a consequence of the WW-relations 
\re{WWG} and \re{WWGt}. 
Let us note, that this cancellation is  not related to the question about behavior
at the points $u=\pm 1$. 
From the calculation one can see that contribution of each diagram is nonsingular
because twist-3 distributions $G(u)$ and $\Gtilde(u)$ enter in a  combination
which nullifies at the end-points.
More  detailed discussion of this question is given in the Appendix.     

The second integral in \re{Mrho} is well defined if tensor gluon SPD $G^T(w,\xi)$
and its first derivative  have no discontinuities at the points 
$x=\pm \xi$. At present the detailed properties of this gluon SPD are 
poorly known. Therefore,
the one possible procedure to get information about behavior at these points is
to study the asymptotic $\mu\to \infty$ limit of SPD evolution. 
The asymptotic expression $G^T_{as}(x,\xi)$ is given by \re{SPDas}. 
Using this expression one obtains that
corresponding  convolution integral is free from singularities.
Assuming that the analytical properties at the points $x=\pm \xi$ 
of the non-asymptotic amplitude are the same, i.e. at least first derivative is 
smooth  function, we conclude  that integral with tensor gluon SPD converges.
Therefore we observe again that the convolution integrals of the twist-3 
amplitude with tensor gluons do not have  divergencies at the end-points.

The tensor structure $S_{\mu\nu}\Delta_\perp^\mu\Delta_\perp^\nu =
\Delta_\perp^\mu\Delta_\perp^\nu-\frac12 g^{\mu\nu}_\perp\Delta^2_\perp$ 
signals a $cos2\phi$
term in the cross section where $\phi$ is angle between lepton and hadron planes.
 The azimuthal dependence can be used   to make a direct extraction
of the double-spin-flip amplitude. The detailed description of the polarization 
density matrix for the $\rho_0$-meson can be found in \ci{Schil}.

Assuming that factorisation takes place we conclude that electroproduction of 
 $\rho$-meson with helicity flip provides a good opportunity to study 
effects due to transverse gluons in a target. 
Concerning to possible higher twist corrections
 we expect that their effect  can not be large. 
Parametrically, a  ratio of twist-3 to  higher twist amplitudes behaves like 
$\frac{f_\rho m_\rho}{Q^2}:\frac{f_\rho m_\rho}{Q^2}\frac{M^2}{Q^2}$, where
$M^2$ is some soft scale, for instance, $\rho$-meson mass or a nucleon mass for
nucleon target. This estimate is based on considerations of possible partonic
contributions with nonzero orbital momentum. At this point one finds
a difference with DVCS helicity flip amplitude. As it was shown in 
\ci{KM} helicity flip amplitude in  DVCS can be  very sensitive to the 
power corrections related to orbital motion of quarks.


\section*{\normalsize \bf Conclusions}

We have calculated twist-3 helicity-flip amplitudes for
the hard meson electroproduction in the Wanzura-Wilczek approximation
on a scalar pion target. We considered two cases corresponding to 
production of C-even tensor meson $f_2$ and  C-odd vector meson $\rho$.
We have found that both amplitudes receive contributions from tensor 
gluon operator. In both cases convolution integrals in the amplitudes 
are free of the endpoint singularities at least for the asymptotical 
approximations of the non-perturbative matrix elements of the tensor 
gluon operator. We conclude therefore that basing  on this observation we 
may suppose that factorisation is not violated for these twist-3 amplitudes.

\section*{\normalsize \bf Acknowlegments.}

The author would like to thank   V.~Braun, 
M.~Polaykov, D.~Ivanov, L.~Mankiewicz, A.~Radyushkin  and O.~Teryaev  
for conversations. The work  was supported by the 
DFG, project No. 920585.

\section*{\normalsize \bf APPENDIX:  Calculation of the $\rho$-meson 
helicity-double-flip amplitude}
\setcounter{equation}{0}
\label{app:a}
\renewcommand{\theequation}{A.\arabic{equation}}
\setcounter{table}{0}
\renewcommand{\thetable}{\Alph{table}}

Here we briefly describe the calculations of the $\rho$-meson amplitude with
helicity flip and discuss contributions of the twist-3 functions $G(u)$ and 
$\Gtilde (u)$. The diagrams contributing to our amplitude are depicted in
Fig.~6. Consider as example calculation of  diagram $D_3$. 
Assuming for simplicity that $N_f=1$ and taking into account definition 
of the amplitude \re{Mrho} we obtain: 
\be{D3in}
D_3^\mu= e(ig)^2\left(\frac{i}{2\pi^2} \right)^2 \int d^4x\int d^4y 
\langle \pi(p'),\rho(Q',\lambda)|
 \bar\psi (x) \hat A^a(x)t^a \frac{\hat x}{x^4}\gamma^\mu
\frac{-\hat y}{y^4}\hat A^b(y)t^b\psi(y)
|\pi(p)\rangle\
\ee
whete we use notation $\hat A= \gamma^\mu A_\mu$ and $t^a$ denotes the 
generators of the SU($N_c$) group which satisfy $\Tr(t^at^b)=\frac12\delta^{ab}$. 
Rewriting this expression in terms of the two matrix elements: 
$\langle p'|A(x)A(y)|p\rangle$  and $\langle Q'|\bar\psi(x)...\psi(y)|0\rangle$
we obtain matrix elements with gluon fields. Remind that we need only
twist-2 tensor projection. Using  axial gauge $n^*A=0$  and definition 
\re{GTSPD} we have \ci{Rad}:
\be{AA}
\langle p'|S_{\alpha\beta} A^\alpha(x)A^\beta(y)|p\rangle =
\frac{\Delta_\perp^{(\alpha\beta)}}{f_\pi^2}
\int_{-1}^{1} dx \frac{G^T(x,\xi)}{(x-\xi+i0)(x+\xi-i0)} 
e^{i(Px)(x-\xi)-i(Py)(x+\xi)}\, ,
\ee
where $S_{\alpha\beta}\Delta_{\perp}^\alpha\Delta_\perp^\beta\equiv
\Delta_\perp^{(\alpha\beta)}$\, .
For the  quark matrix element it is convenient to use Fiertz identities
\footnote{We define Dirac matrix  $\gamma_5$ as 
$\gamma_5=i\gamma^0\gamma^1\gamma^2\gamma^3$ and
$\Tr(\gamma_5\gamma_{\mu_1}\gamma_{\mu_2}\gamma_{\mu_3}\gamma_{\mu_4})=
4i\eps_{\mu_1\mu_2\mu_3\mu_4}$}:
\be{Fid}
\langle Q'|\bar\psi(x)\Gamma\psi(y)|0\rangle=\frac14 \Tr\{\gamma_\sigma\Gamma \}
\langle Q'|\bar\psi(x)\gamma^\sigma\psi(y)|0\rangle -
\frac14 \Tr\{\gamma_\sigma\gamma_5\Gamma \}
\langle Q'|\bar\psi(x)\gamma^\sigma\gamma_5\psi(y)|0\rangle\, ,
\ee   
where $\Gamma$ is generic Dirac matrix structure. Perfoming these transformations 
and substituting gluon matrix element in \re{D3in} we arrive at:
\be{D3out}
D_3^\mu&=& -\frac{eg^2}{4\pi^4}\frac{1}{2N_c}\frac{1}{f_\pi^2}
\int_{-1}^{1} dx \frac{G^T(x,\xi)}{(x-\xi+i0)(x+\xi-i0)}
\int d^4x\int d^4y e^{i(Px)(x-\xi)-i(Py)(x+\xi)}
\frac{\Delta_\perp^{(\alpha\beta)}}{x^4y^4}
\nonumber \\[4mm]
&&\times \left\{ 
\frac14\Tr\{\gamma_\sigma\gamma_\alpha\hat x\gamma^\mu\hat y \gamma_\beta\}
\langle Q'|\bar\psi(x)\gamma^\sigma\psi(y)|0\rangle-
\frac14\Tr\{\gamma_\sigma\gamma_5\gamma_\alpha\hat x\gamma^\mu\hat y \gamma_\beta\}
\langle Q'|\bar\psi(x)\gamma^\sigma\gamma_5\psi(y)|0\rangle
\right\}. \nonumber
\ee
At the next step one has to substitude into \re{D3out} twist-3 expressions \re{Vrho}
and \re{Arho} for the quark matrix elements and perfom coordinate-space integrations.
 The calculation of the other diagrams can be done
using the similar approach. The final results can be written as follows:
\be{D1}
D_1^\mu&=&\frac{\Delta_\perp^{(\mu \nu)}}{Q^2}e^*_\nu C 
\int_{-1}^{1}dx \frac{(-1)G^T(x,\xi)}{(x-\xi+i0)(x+\xi-i0)^2}
\int_{-1}^{1}\frac{du}{1-u^2} \left[ 
G(u)-\frac u2 \Gtilde(u)-\phi(u)
\right]\, , 
\\[4mm]
\label{D2}
D_2^\mu&=&\frac{\Delta_\perp^{(\mu \nu)}}{Q^2}e^*_\nu C 
\int_{-1}^{1}dx \frac{G^T(x,\xi)}{(x-\xi+i0)^2(x+\xi-i0)}
 \int_{-1}^{1}\frac{du}{1-u^2} \left[ 
G(u)-\frac u2 \Gtilde(u)-\phi(u)
\right]\, ,
\\[4mm]
\label{D3}
D_3^\mu&=&\frac{\Delta_\perp^{(\mu \nu)}}{Q^2}e^*_\nu C 
\int_{-1}^{1}dx \frac{G^T(x,\xi)}{(x-\xi+i0)^2(x+\xi-i0)^2}
(-2\xi) \int_{-1}^{1} \frac{du}{1-u^2}\left[ 
G(u)-\frac u2 \Gtilde(u)
\right]\, , 
\ee
where the common factor 
$C=\frac{f_\rho m_\rho}{f_\pi^2}(4\pi e\alpha_S)\frac{-2\xi}{N_c}$.  
 We have also used the properties of the WW-contributions 
 \re{WWG} and \re{WWGt} to rewrite the contributions $D_1^\mu$ and 
$D_2^\mu$  in the form presented in \re{D1} and \re{D2}. Using  
\re{WWG} and \re{WWGt} one finds that
functions $G(u)$ and $\Gtilde(u)$ do not equal zero at the end points:
\be{bound}
G(\pm 1)=\frac12\int_{-1}^1 du\frac{\phi(u)}{1-u}\, ,\, \,
\Gtilde(\pm 1)=\pm\int_{-1}^1 du\frac{\phi(u)}{1-u}\, ,
\ee 
In our case they enter into the expressions \re{D1}--\re{D3} in the
combination $G(u)-\frac u2 \Gtilde(u)$ which nullifies at 
the points $u=\pm1$. The latter is very important
for convergence of the corresponding integrals in 
\re{D1}--\re{D3}. However, in the sum of the all  contributions 
these WW-terms cancel:
\be{sum}
D_1^\mu+D_2^\mu+D_3^\mu= 
\frac{\Delta_\perp^{(\mu \nu)}}{Q^2}e^*_\nu C (-2\xi) 
\int_{-1}^{1}dx 
\frac{G^T(x,\xi)}{(x-\xi+i0)^2(x+\xi-i0)^2}
\int_{-1}^{1}du \frac{\phi(u)}{1-u^2}\,  . 
\ee
As we note in the main body of the paper it may be related to a property
of the WW-approximation. To get final answer \re{Mrho} from \re{sum}  
one has to take into account the
flavor structure which reduces to  multiplication of \re{sum} on the factor 
$\frac12(e_u-e_d)=\frac12$.

\newpage

\begin{figure}[p]
\unitlength1mm
\begin{center}
\hspace{0cm}
\insertfig{6}{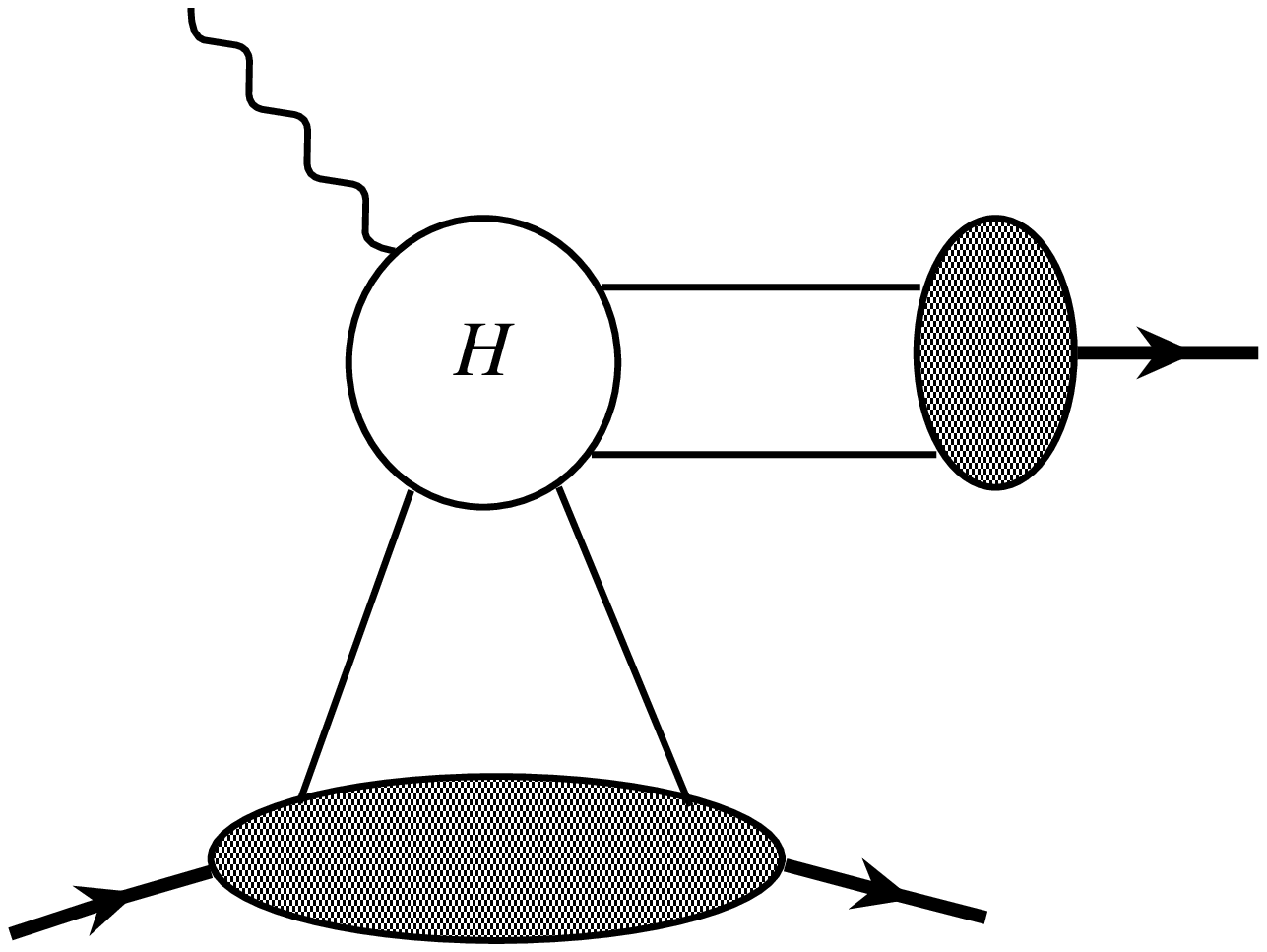}
\end{center}
\vspace{-0.5cm}
\caption[dummy]{\small OPE-like regime for the hard electroproduction 
of meson.  
\label{hard}}
\end{figure}

\begin{figure}[p]
\unitlength1mm
\begin{center}
\hspace{0cm}
\insertfig{6}{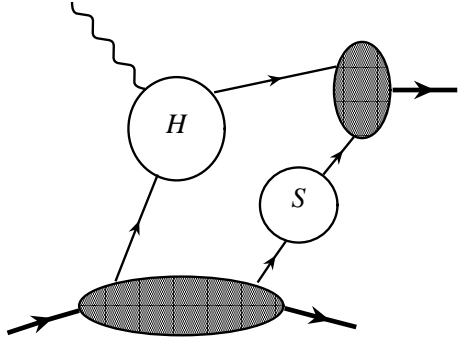}
\end{center}
\vspace{-0.5cm}
\caption[dummy]{\small Endpoint regime.
\label{hardS}}
\end{figure}

\newpage

\begin{figure}[p]
\unitlength1mm
\begin{center}
\hspace{0cm}
\insertfig{6}{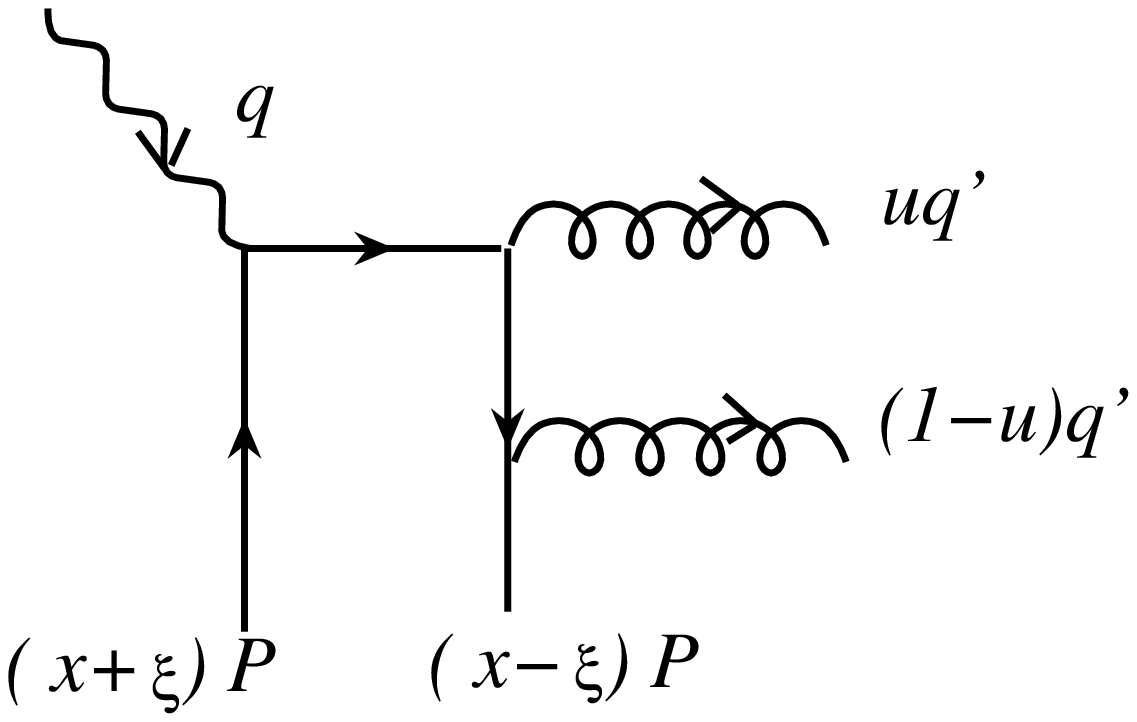}
\hskip0.5cm
\insertfig{8}{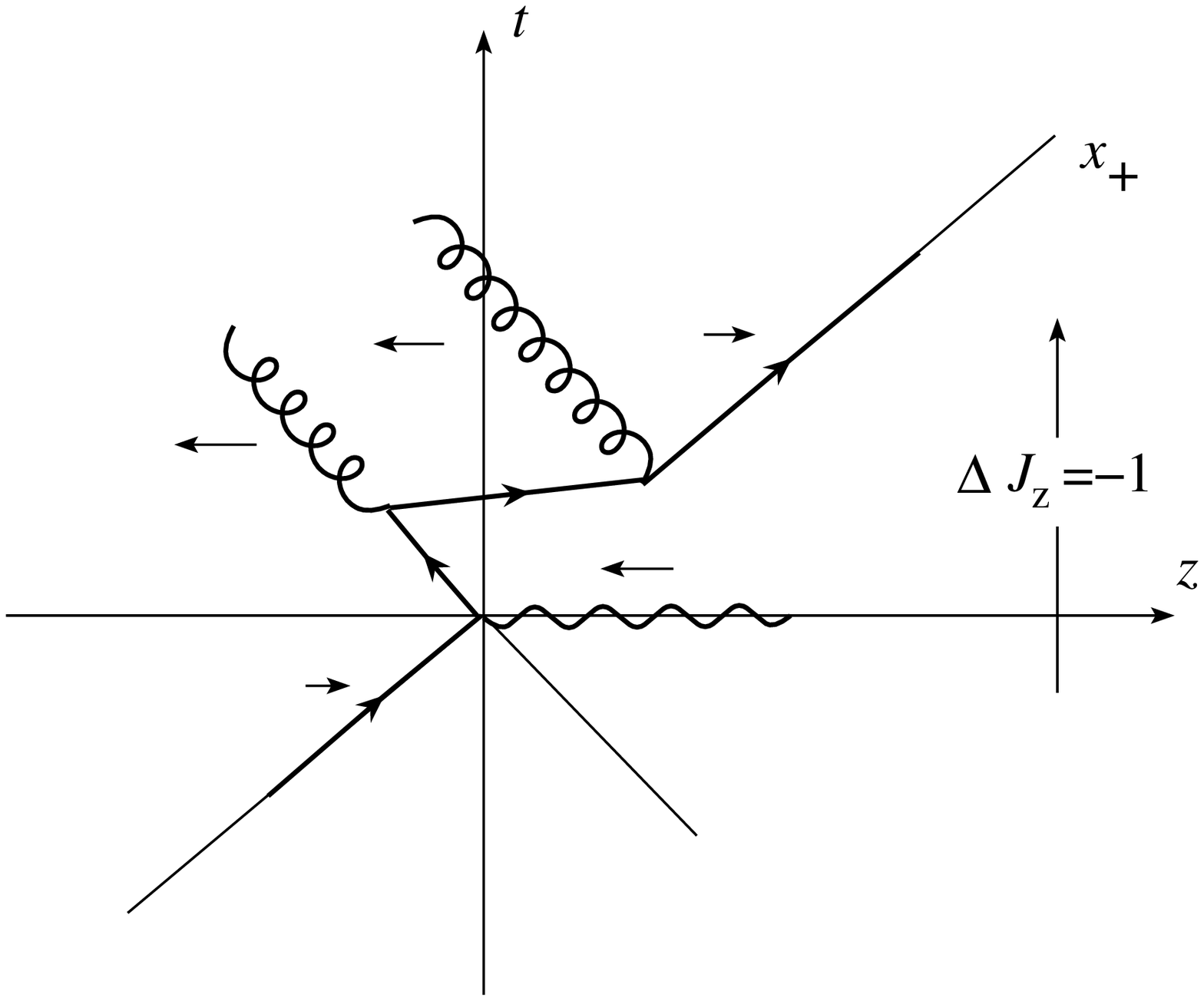}
\end{center}
\vspace{-0.5cm}
\caption[dummy]{\small The partonic subprocess for the production of the
$f_2$-meson  by tensor gluons (left) and its space time
development (right). The arrows above each partonic line 
denote the projection of the spin of 
the particle on $z$-axis.  
\label{spaceG}}
\end{figure}

\begin{figure}[p]
\unitlength1mm
\begin{center}
\hspace{0cm}
\insertfig{6}{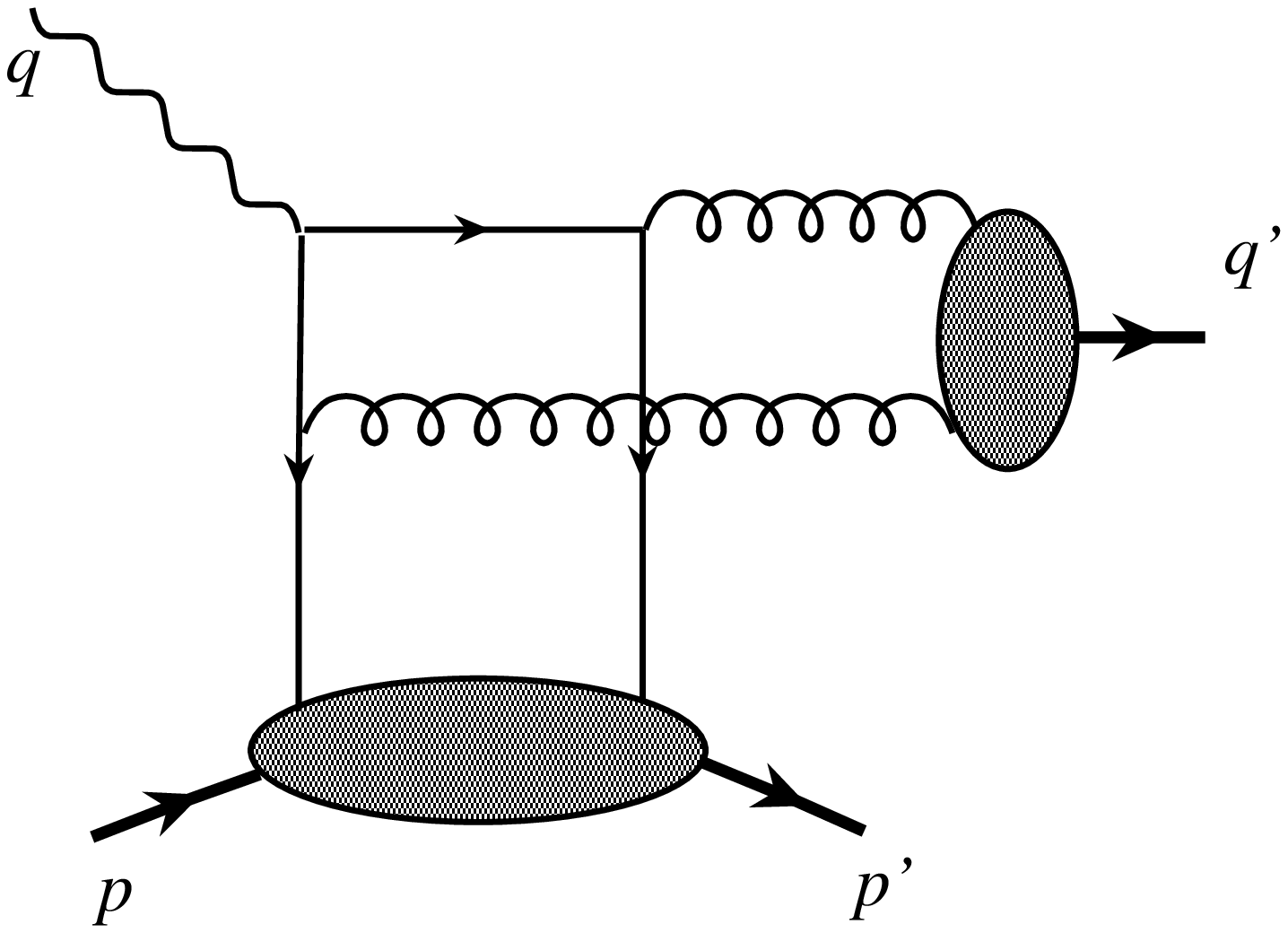}
\hskip0.5cm
\insertfig{6}{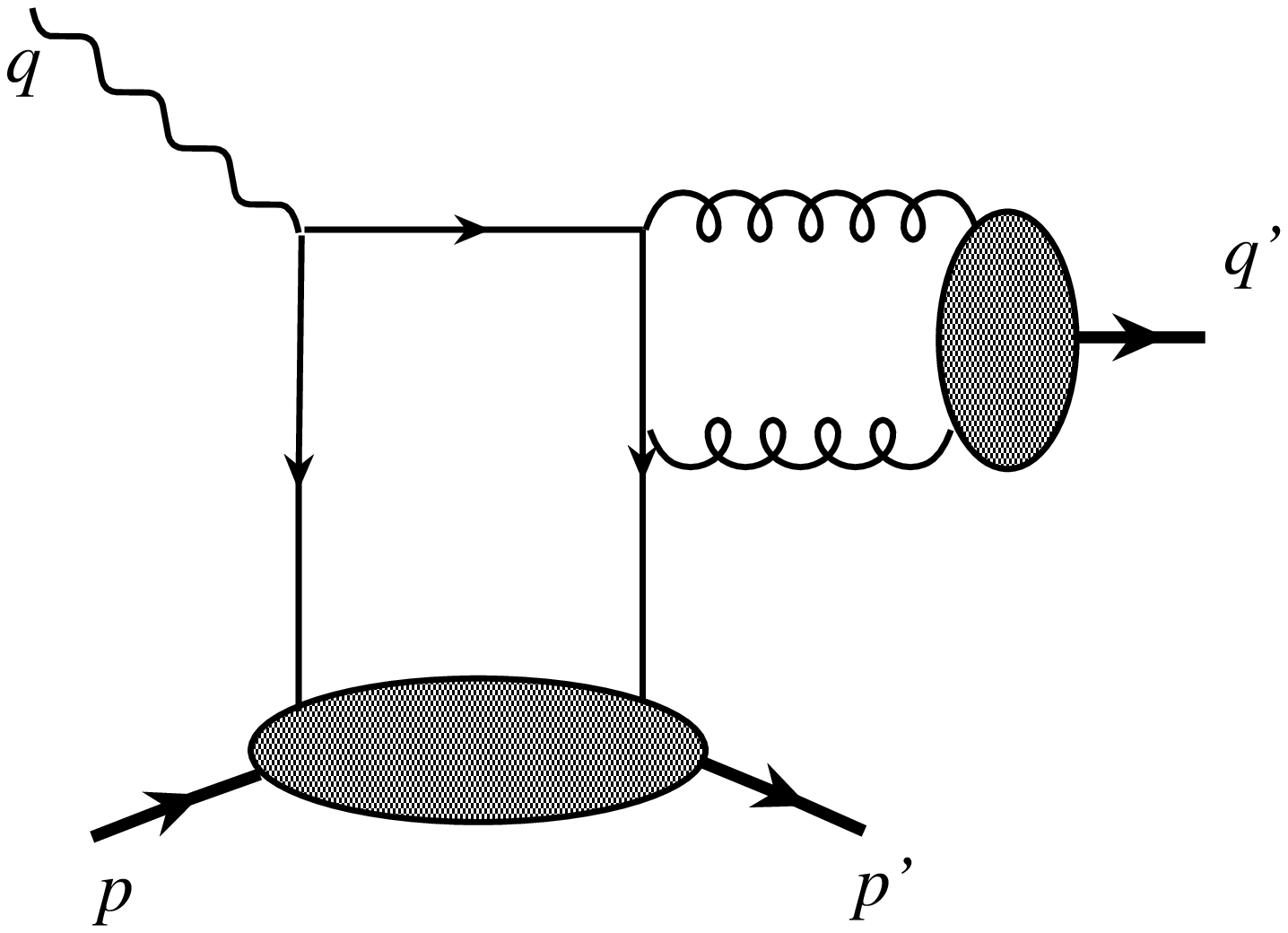}
\end{center}
\vspace{-0.5cm}
\caption[dummy]{\small 
Typical diagrams for electroproduction of $f_2$ with gluon 
distribution amplitudes.
\label{diag34}}
\end{figure}

\begin{figure}[p]
\unitlength1mm
\begin{center}
\hskip0.5cm
\insertfig{6}{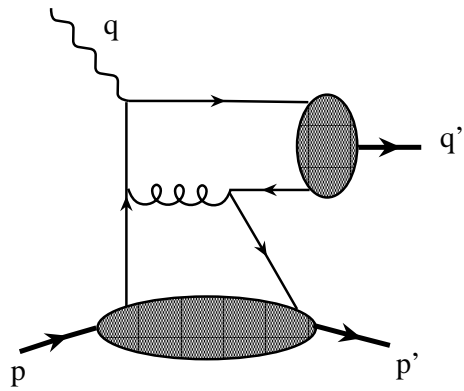}
\end{center}
\vspace{-0.5cm}
\caption[dummy]{\small 
Typical diagrams for electroproduction of $f_2$ with quark 
distribution amplitude.
\label{diag6}}
\end{figure}

\newpage

\begin{figure}[p]
\unitlength1mm
\begin{center}
\hspace{0cm}
\insertfig{6}{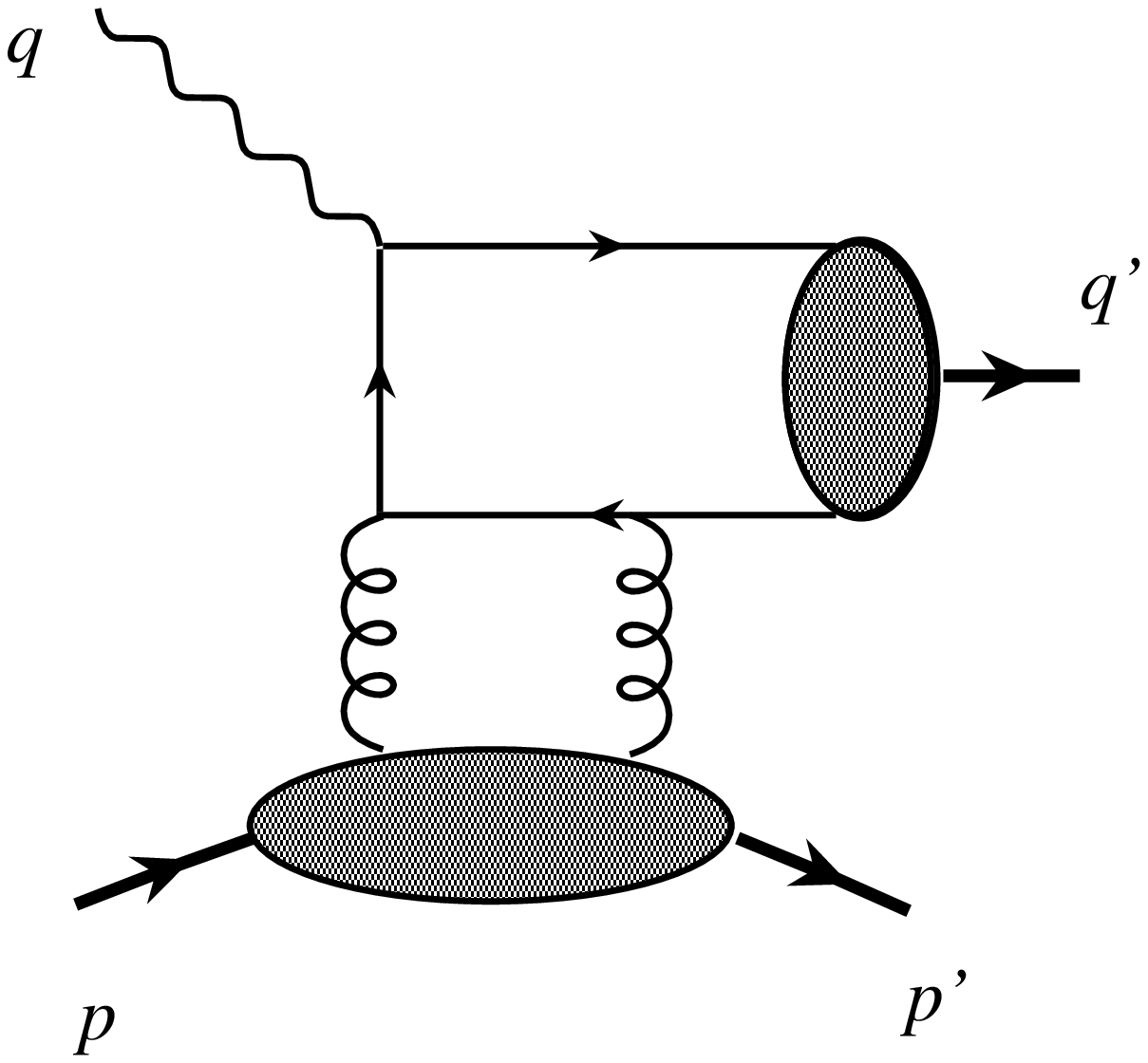}
\hskip0.5cm
\insertfig{6}{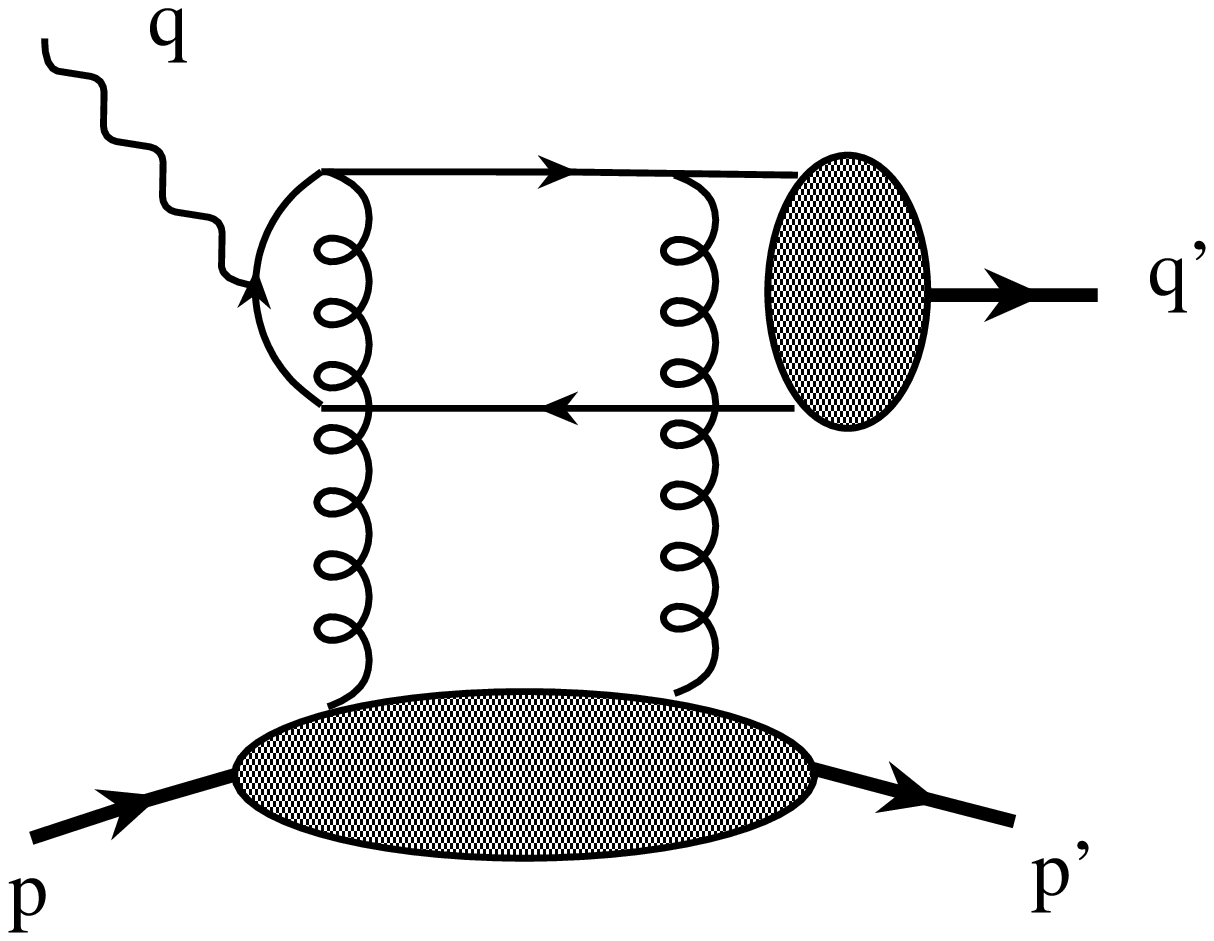}
\insertfig{6}{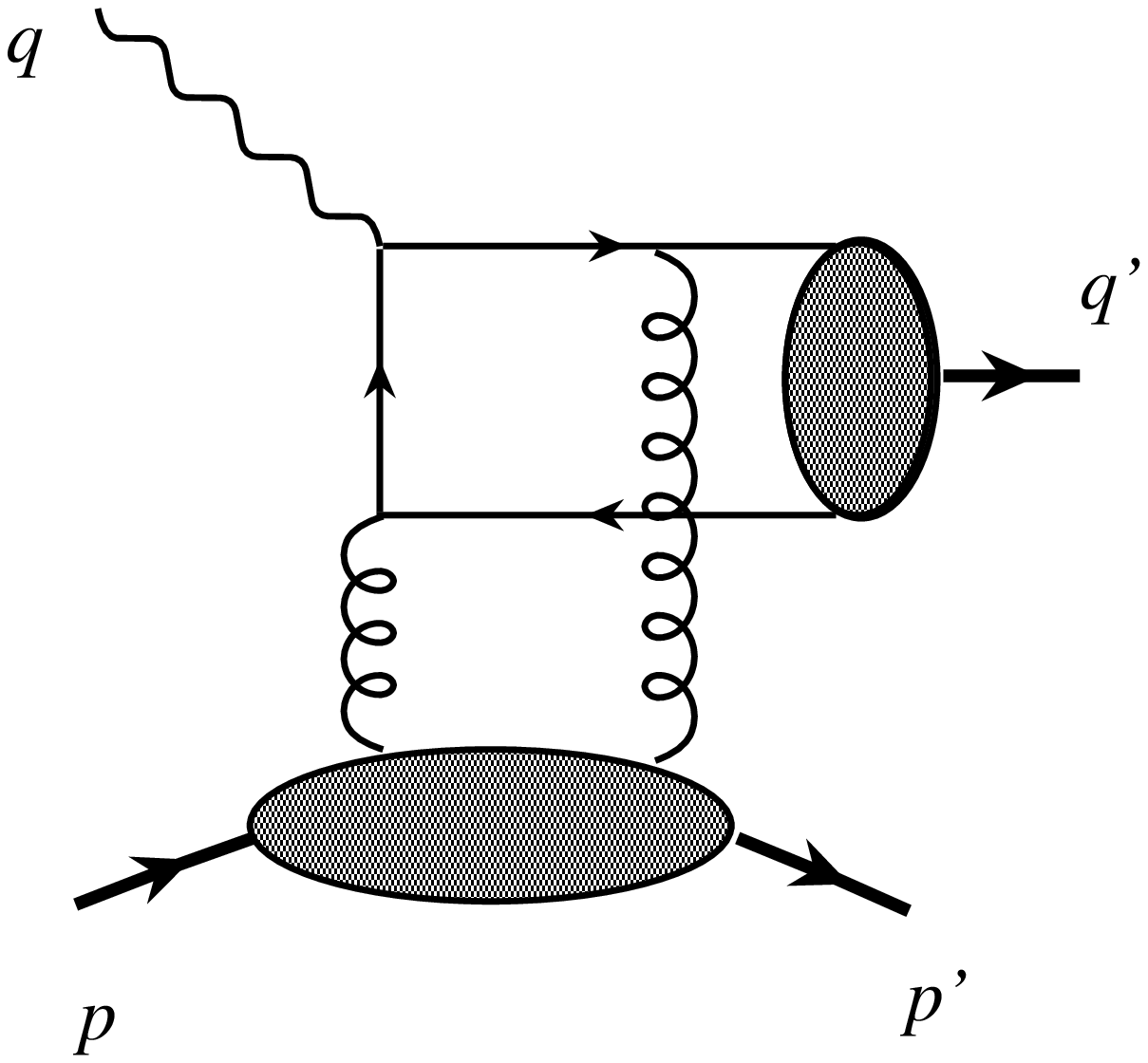}
\end{center}
\vspace{-0.5cm}
\caption[dummy]{\small Diagrams for the electroproduction of
$\rho$-meson with double-helicity-flip. 
Their contributions are denoted as D1, D2 and D3 respectively.
\label{diag125}}
\end{figure}

\begin{figure}[p]
\unitlength1mm
\begin{center}
\hspace{0cm}
\insertfig{6}{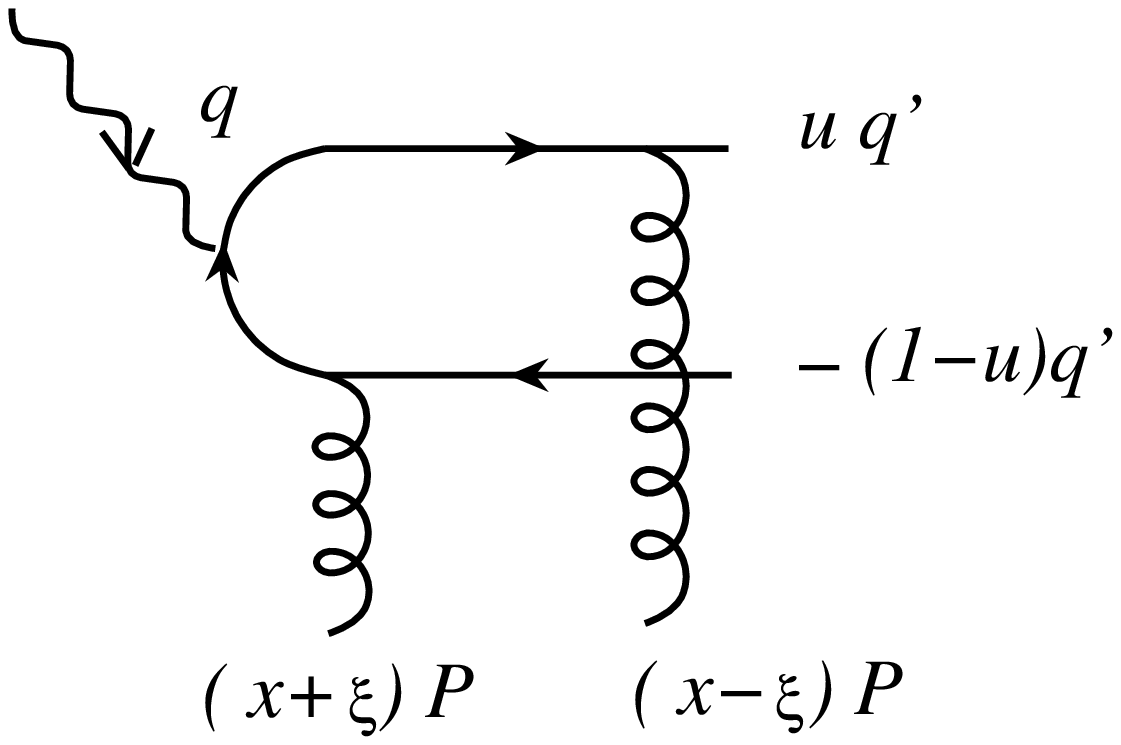}
\hskip0.5cm
\insertfig{8}{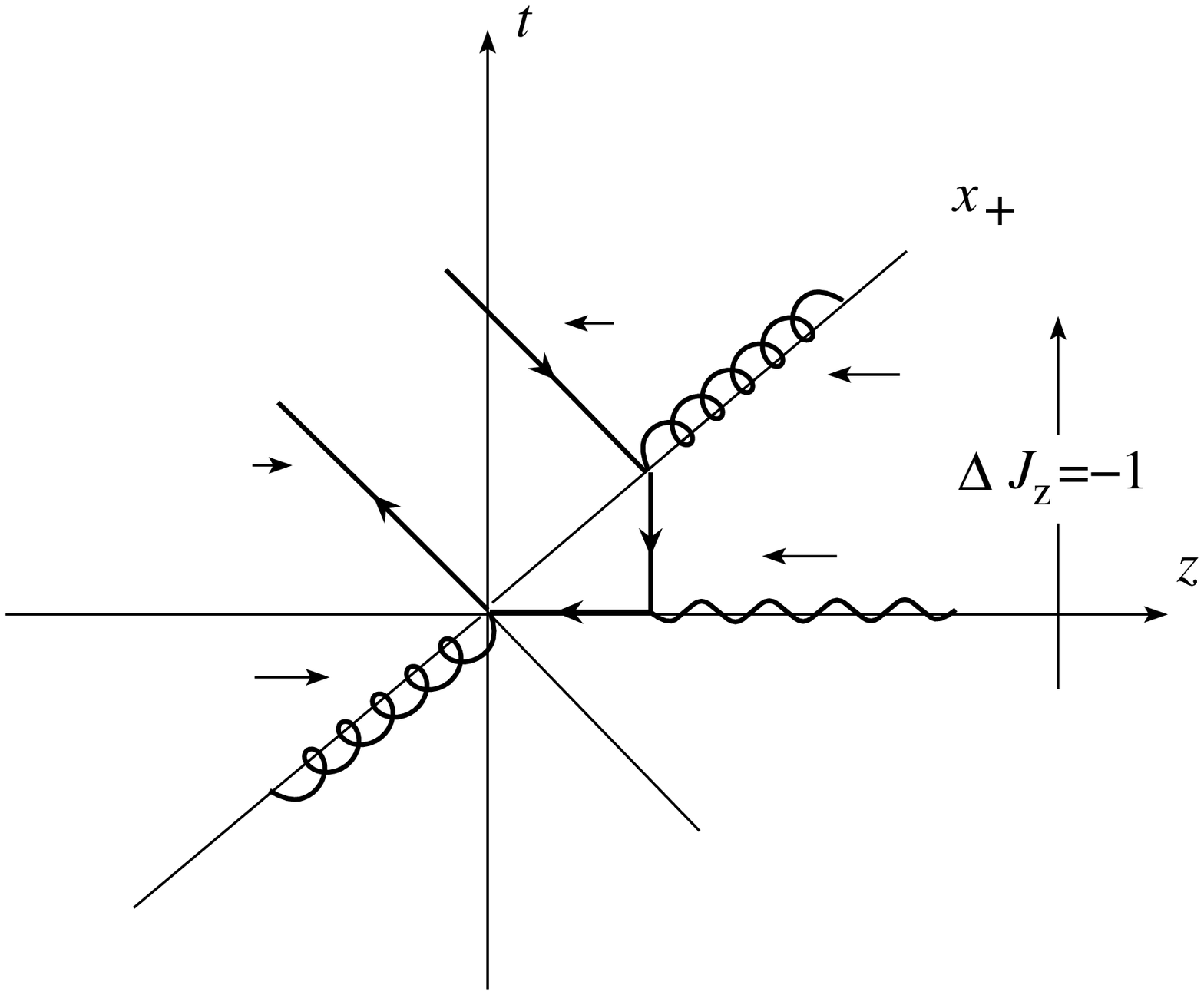}
\end{center}
\vspace{-0.5cm}
\caption[dummy]{\small 
 The partonic subprocess for the double-helicity-flip amplitude of the
$\rho$-meson  (left) and its space time
development (right). The arrows above each partonic line  denote the 
projection of the spin of the particle on $z$-axis.
\label{spGT}}
\end{figure}

\end{document}